\documentclass[12pt]{article}
\usepackage[T1]{fontenc} 
\usepackage[utf8]{inputenx}
\usepackage{babel} 
\usepackage{url}
\usepackage{geometry}
\usepackage{amsmath}
\usepackage{amsfonts}
\usepackage{amstext}
\usepackage{amssymb}
\usepackage{amsthm}
\usepackage{amscd}
\usepackage{mathrsfs}
\usepackage{xcolor}
\definecolor{myurlcolor}{rgb}{0,0,0.4}
\definecolor{mycitecolor}{rgb}{0,0.5,0}
\definecolor{myrefcolor}{rgb}{0.5,0,0}
\usepackage{graphicx}
\usepackage{tikz}
\usepackage{tikz-cd}
\usepackage{etaremune}
\usepackage{latexsym}
\usepackage{braket}
\usepackage{soul}
\setstcolor{blue}
\usepackage{comment}
\usepackage{etoolbox}
\usepackage{makeidx}
\usepackage{dsfont}
\usepackage{enumitem} 
\usepackage{caption}
\usepackage{lmodern}
\usepackage{textcomp}
\usepackage{totcount}
\usepackage{blindtext}

\usepackage[pagebackref,draft=false]{hyperref}
\hypersetup{colorlinks,linkcolor=myrefcolor,citecolor=mycitecolor,urlcolor=myurlcolor}

\usepackage[capitalize]{cleveref}

\usepackage{autonum}

\newtheorem*{proof*}{Proof}

\title{Towards a category-theoretic foundation of Classical and Quantum Information Geometry}

\author{F. M. Ciaglia$^{1,5}$ \href{https://orcid.org/0000-0002-8987-1181}{\includegraphics[scale=0.7]{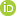}}, F. Di Cosmo$^{2,3,6}$ \href{https://orcid.org/0000-0003-0256-5913}{\includegraphics[scale=0.7]{ORCID.png}}, L. González-Bravo$^{3,4,7}$ \href{https://orcid.org/0000-0002-4382-7978}{\includegraphics[scale=0.7]{ORCID.png}}}

\begin{document}

\maketitle 
\vspace{-0.5cm}
\noindent
{\footnotesize $^{1}$  Universidad Carlos III de Madrid, ROR: \href{https://ror.org/03ths8210}{03ths8210}, Departamento de Matemáticas, Avenida de la Universidad, 30 (edificio Sabatini), 28911 Leganés (Madrid), España.}  \\
{\footnotesize $^{2}$ Universidad de Alcalá, ROR: \href{https://ror.org/04pmn0e78}{04pmn0e78}, Departamento de Física y Matemáticas, Ctra Madrid-Barcelona, km.33, 600, 28805 Alcalá de Henares, Madrid, España.\\
{\footnotesize $^{3}$ Instituto de Ciencias Matem\'{a}ticas ICMAT (CSIC-UAM-UC3M-UCM), ROR: \href{https://ror.org/05e9bn444}{05e9bn444}, Campus de Cantoblanco UAM, Calle Nicolás Cabrera, 13-15, 28049 Madrid, España.} \\}
{\footnotesize $^{4}$ Universidad Complutense de Madrid, ROR: \href{ https://ror.org/02p0gd045}{02p0gd045}. Departamento de Álgebra, Geometría y Topología, Facultad de Ciencias Matemáticas, Pl. de las Ciencias, 3, Moncloa-Aravaca, 28040 Madrid, España}

\bigskip
\noindent
{\footnotesize $^{5}$\texttt{fciaglia[at]math.uc3m.es}  $^{6}$\texttt{fcosmo[at]math.uc3m.es}   $^{7}$\texttt{lauraego[at]ucm.es}}

\begin{abstract}
We introduce the category $\mathsf{NCP}$, whose objects are pairs of W$^\ast$-algebras and normal states and whose morphisms are state-preserving unital completely positive (CPU) maps, as a common stage for classical and quantum information geometry, and  we formulate two  results that will appear in forthcoming works. 
%%%%%%%%%%%%%%%%%%%%%%%%%%%%%%%
 
First, we recast the problem of classifying admissible Riemannian geometries on classical and quantum statistical models in terms of functors $\mathfrak{C}:\mathsf{NCP}\to\mathsf{Hilb}$.
%%%%%%%%%%%%%%%%%%%%%
These functors provide a generalization of classical statistical covariance, and we call them \textit{fields of covariances}. A prominent example being the so-called GNS functor arising from the Gelfand-Naimark-Segal (GNS) construction. 
%%%%%%%%%%%%%%%%%%%%%%%%%%%%%%%%
The classification of fields of covariances on $\mathsf{NCP}$ entails both Čencov's uniqueness of the Fisher-Rao metric tensor and Petz's classification of monotone quantum metric tensors as particular cases. 
%%%%%%%%%%%%%%%%%%%%%%

Then, we show how classical and quantum statistical models can be realized as subcategories of $\mathsf{NCP}$ in a way that takes into account symmetries. 
%%%%%%%%%%%%%%%%%
In this setting, the fields of covariances determine Riemannian metric tensors on the model that reduce
to the Fisher-Rao, Fubini-Study, and Bures-Helstrom metric tensor in particular cases.
\end{abstract}

{\footnotesize
\tableofcontents
}

\section{Introduction}
In statistics, the Fisher-Rao metric tensor $G_{FR}$ provides a lower bound for the covariance of an unbiased estimator through the so-called Cramér-Rao inequality \cite{Cramer-1946}.
%%%%%%%%%%%%%%%%%%%%%%%%%%%%%%%%%%%%
Since this inequality applies broadly across a wide range of statistical models, it raises a natural question: why does the ``same'' metric tensor consistently appear in such diverse settings?
%%%%%%%%%%%%%%%%%%%%%%%%%%%%

In his work \cite{Cencov-1982}, Čencov investigated this question providing a detailed answer in the case of finite probability spaces $\mathcal{X}_{n+1}$.
%%%%%%%%%%%%%%%%%%%%%%%%%%%%%%%%%%%
Specifically, Čencov built the category $\mathsf{CAPF}$ whose objects are the smooth manifolds $\Delta_n$ given by the open interior of probability simplexes $\overline{\Delta}_n$, and whose morphisms are Markov maps \cite{Cencov-1982}.  
%%%%%%%%%%%%%%%%%%%%%%%%%%%%%%%%%%%%%%%%%%%%%%%%%%%%%%%
Within this framework, Čencov proved that the Fisher-Rao metric tensor is the only metric tensor on objects of $\mathsf{CAPF}$ which is invariant under congruent embeddings (that is, the smooth maps determined by those Markov maps admitting a left inverse which is also a Markov map).
%%%%%%%%%%%%%%%%%%%%%%%%%%%%%%%%%%%%%%%%%%%%%%%%%%%%%%%

Čencov's original insights naturally extend to the quantum setting, a direction that he and Morozova explored in \cite{C-M-1987,C-M-1991}. 
%%%%%%%%%%%%%%%%%%%%%%%%%%%%%%%%%
In the finite-dimensional quantum case, we can define the category $\mathsf{cQS}$ of quantum states, whose objects are the manifolds of faithful quantum states $\mathscr{S}_f(\mathcal{H})$ on the Hilbert space $\mathcal{H}$ of the system, and whose morphisms are determined by completely-positive, trace preserving (fCPTP) maps\footnote{These maps are the quantum counterparts of classical Markov morphisms \cite{Choi-1975,N-C-2011}.}.
%%%%%%%%%%%%%%%%%%%%%%%%%%%%%%%%%%%%%
In this context, the requirement of invariance with respect to congruent embeddings characterizing the Fisher-Rao metric tensor in the classical case is replaced by the stronger requirement that quantum morphisms are contractions (the so-called \textit{monotonicity property}).
%%%%%%%%%%%%%%%%%%%%%%
Despite a stronger requirement, the quantum case presents an infinite number of Riemannian metric tensors satisfying the monotonicity property, as firstly discussed by Čencov and Morozowa \cite{C-M-1991}, and later rigorously proved by Petz \cite{Petz-1996}, who classified all these Riemannian metric tensors in terms of suitable operator monotone functions.
%%%%%%%%%%%%%%%%%%%%

Čencov's and Petz's classifications of metric tensors focus on the manifolds of faithful probability vectors and faithful quantum states in finite dimensions, respectively.
%%%%%%%%%%%%%%%%%%%%%%%%%%%%%%%%%%%%%%%%%%%%%%%%
In particular, Čencov's work leaves outside many important statistical models which have continuous outcome spaces (\textit{e.g.}, multivariate normal distributions), while Petz's classification excludes all models on infinite-dimensional algebras, and all models of non-faithful states (\textit{e.g.}, pure states\footnote{In \cite{P-S-1996}, Petz and Sudar discuss an \textit{ad hoc} procedure to recover the Fubini-Study metric tensor on pure quantum states as a suitable limit of the Riemannian metric tensors on faithful states.}, which are highly relevant in quantum mechanics).
%%%%%%%%%%%%%%%%%%%%%%%%%%%%%%%%%%%%%%%%%%%%%%%%

Several efforts have been made to extend Čencov's theorem to the case of continuous outcome spaces, as demonstrated in works such as \cite{Michor-1980,A-J-L-S-2017,Fujiwara-2023,P-S-1995}.
%%%%%%%%%%%%%%%%%%%%%%%%%%%%%%
However, none of these generalizations engage with Čencov's original categorical framework. %and they still depend on the specifics of the underlying statistical model employed. 
%%%%%%%%%%%%%%%%%%%%%%%%%%%%%%%%%%%%%
Moreover, while Čencov's original work naturally extends to the quantum setting, the mathematical techniques employed in these generalizations are not well-suited to be adapted to the quantum domain.
%%%%%%%%%%%%%%%
Finally, to the best of the authors' knowledge, no attempt to generalize Petz's theorem to infinite-dimensions has been undertaken. 
%%%%%%%%%%%%%%%%%%%%%%%%%%%%%%%%%%%%%

These challenges point to the need for a more flexible and structurally unified approach.
%%%%%%%%%%%%%%%%%%%%%%%%%%%%%%%%%%%%%%%%%%%%%%%%%%%
Operator algebras offer a natural mathematical setting in which the structural similarities between classical and quantum theories can coexist, providing a common language for describing and analyzing their probabilistic aspects.
%%%%%%%%%%%%%%%%%%%%%%%%%%%%%%%%%%%%%%%%%%%%%%%%%%%
This unifying perspective is particularly appealing for the development of a general framework for information geometry, one that encompasses both classical and quantum settings.
%%%%%%%%%%%%%%%%%%%%%%%%%%%%%%%%%%%%%%%%%%%%%
The key idea is that from a functional-analytic perspective, both (dominated) probability distributions and quantum states (or density operators) can be understood as distinct realizations of the notion of a normal state—on an Abelian W$^*$-algebra in the classical case, and on a non-Abelian W$^*$-algebra in the quantum case.\footnote{For all the details on operator algebras that are used but not discussed in this work, we refer the reader to \cite{Blackadar-2006,takesaki-2002}.}
%%%%%%%%%%%%%%%%%%%%%%%%%%%%%%%%%%%%%%%%%%%%%
This unified framework is particularly motivated by the presence of probabilistic structures in quantum mechanics that closely mirror their classical counterparts. 
%%%%%%%%%%%%%%%%%%%%%%%%%%%%%%%%%%%%%%%%%%%%%%%%
For example, the Born rule assigns probabilities to measurement outcomes in a way that parallels classical probability theory, while positive operator-valued measures (POVMs) generalize quantum measurements in a manner that resembles classical probabilistic processes. 
%%%%%%%%%%%%%%%%%%%%%%%%%%%%%%%%%%%%%%%%%%%%%%%%%
Other examples include local operations and classical communication (LOCC) which directly combine local quantum operations with classical coordination and quantum tomography which adapts classical statistical inference techniques to the quantum setting.
%%%%%%%%%%%%%%%%%%%%%%%%%%%%%%%%%%%%%%%%%%%%%%%%%%%%

Within the operator algebraic approach, there is a plethora of interesting results that highlight deep connections between algebraic structures and information geometry \cite{C-J-S-2020,Jencova-2003,Jencova-2006,Jencova-2023,C-DN-J-S-2023}.
%%%%%%%%%%%%%%%%%%%%%%%%%%%%%%%%%
%Examples include the derivation of metrics such as the Fisher–Rao, Fubini–Study, and Bures–Helstrom from the Jordan structure of finite-dimensional C*-algebras \cite{C-J-S-2020}; the extension of key information-geometric concepts to the setting of von Neumann algebras \cite{Jencova-2003}; the rigorous construction of infinite-dimensional differential manifold structures on the space of faithful states of a quantum system represented by a von Neumann algebra \cite{Jencova-2006,Jencova-2023}; and the generalization of parametric statistical models by means of normal positive linear functionals and W*-algebras \cite{C-DN-J-S-2023}.
Moreover, in \cite{C-DC-GB-2023} the formulation of classical
and quantum information geometry in terms of (normal) states on finite-dimensional W*-algebras allows for a framework that unifies Čencov’s and Petz’s problems.
%%%%%%%%%%%%%%%%%%%%%%%%%%%%%%%%%%%
However, within this approach a counterpart to the problem originally proposed by Čencov that scales to infinite dimensions and unifies all classical and quantum systems has not yet been formulated.
%%%%%%%%%%%%%%%%%%%%%%%%%%%%%%%%%%%%

In this work we would like to combine the approach of operator algebras and the initial categorical ideas from Čencov to propose a formalism compatible with these requirements. 
%%%%%%%%%%%%%%%%%%%%%%%%%%%%%%%%%%%%%%%%%%
Additionally, in this formalism we would like to implement a way in which the geometry of the models is derived from the ``geometry'' of the ambient space.
%%%%%%%%%%%%%%%%%%%%%%%%%%%%%%%%%%%%%%%%%%
However, this implementation needs to be done in an appropriate suitable manner since as we know, the set of states of an infinite-dimensional W*-algebra is not a smooth manifold\footnote{This is already true in finite-dimensions where we found that such sets are given by the $n$-simplex in the classical case and the set of quantum states in the quantum one.}. 
%%%%%%%%%%%%%%%%%%%%%%%%%%%%%%%%%%%%%%%%%%%
Thus, to get such an implementation we will abandon the idea of manifolds and we will replace it by the notion of a suitable category. 
%%%%%%%%%%%%%%%%%%%%%%%%%%%%%%%%%%%%%%%%%
Of course, in this categorical setting, analogue structures to those of the manifold-oriented approach to information geometry will need to be defined. 
%%%%%%%%%%%%%%%%%%%%%%%%%%%%%%%%%%%%%%%%%
Specifically, building on Čencov’s categorical approach \cite{C-M-1987} and the realization that quantum and classical information geometry can be seen, inspired by Voiculescu's free probabilities \cite{V-D-N-1992}, as different faces of the same non-commutative object, we introduce the category of non-commutative probabilities ($\mathsf{NCP}$) whose objects are given by pairs $(\mathscr{A},\rho)$ where, $\mathscr{A}$ is a W*-algebra and $\rho$ is a normal state and whose morphism are given by state-preserving CPU maps, i.e., a morphism $\Phi: (\mathscr{A},\rho)\to(\mathscr{B},\sigma)$ exists when there is a CPU map $\phi\colon \mathscr{B}\rightarrow \mathscr{A}$ such that $\phi^*(\rho) = \sigma$.
%%%%%%%%%%%%%%%%%%%%%%%%%%%%%%%%%%%%%%%%

The category $\mathsf{NCP}$ can be seen as a sort of environment containing any statistical model, whether classical or quantum.
%%%%%%%%%%%%%%%%%%%%%%%%%%%%%%%%%%%%%%%
Indeed, as we will show, any statistical model can be recovered—at least in some sense—within the language of $\mathsf{NCP}$.
%%%%%%%%%%%%%%%%%%%%%%%%%%%%%%%%%%%%%%
This framework thus provides a categorical setting in which statistical models themselves can be studied.
%%%%%%%%%%%%%%%%%%%%%%%%%%%%%%%%%%%%%%%%%%
Within the categorical context provided by $\mathsf{NCP}$ we aim not only to recover Čencov's and Morozova's categorical perspective but also to develop a suitable categorical counterpart to the problem of invariant geometries in the category CAPF originally posed by Čencov.
%%%%%%%%%%%%%%%%%%%%%%%%%%%%%%%%%%%%%%%
Our proposed Čencov-like problem within $\mathsf{NCP}$ does not depend on the standard notion of a statistical model.
%%%%%%%%%%%%%%%%%%%%%%%%%%%%%%%%%%%%%%
This opens the door to investigating whether the invariant “geometries” can be regarded as model-independent features of information geometry—that is, whether there exists a ``geometry'' on the ambient space $\mathsf{NCP}$ that naturally ``projects'' to the geometries found within individual statistical models.
%%%%%%%%%%%%%%%%%%%%%%%%%%%%%%%%%%%%%%%

This work is structured as follows. 
%%%%%%%%%%%%%%%%%%%%%%%%%%%%%%%%%%%%%%%%%
In Section \ref{sec: section on covariances}, we introduce the novel concept of field of covariances and explain how this concept allows to develop a categorical counterpart to the problem of invariant geometries initiated by Čencov. 
%%%%%%%%%%%%%%%%%%%%%%%%%%%%%%%%%%%%%%%%%
In Section \ref{sec: section on statistical models}, we demonstrate how statistical models can be recovered as subcategories of $\mathsf{NCP}$.

\section{From metric tensors to covariances}\label{sec: section on covariances}

%A classification problem in the spirit of Čencov relies on the presence of geometrical objects such as manifolds and Riemanian metric tensors.
%%%%%%%%%%%%%%%%%%%%%%%%%%%%%%%%%%%%%%%%%
%However, within the categorical framework of $\mathsf{NCP}$ such geometrical structures are absent, as the objects of $\mathsf{NCP}$, unlike those in the categories $\mathsf{CAPF}$ and $\mathsf{cQS}$, are not smooth manifolds. 
%%%%%%%%%%%%%%%%%%%%%%%%%%%%%%%%%%%%%%%%%%
%Thus, a suitable reformulation of Čencov's problem, adapted to the categorical context of $\mathsf{NCP}$ is necessary.  
%%%%%%%%%%%%%%%%%%%%%%%%%%%%%%%%%%%%
%In such a reformulation, we will need to define analogous structures to replace those of the manifold-oriented approach.
%%%%%%%%%%%%%%%%%%%%%%%%%%%%%%%%%%%%%
%To build intuition about these analogues, drawing a parallel between $\mathsf{NCP}$ and the manifold-oriented perspective of information geometry may be helpful. 
%%%%%%%%%%%%%%%%%%%%%%%%%%%%%%%%%%%%%

Intuitively speaking, we can think of $\mathsf{NCP}$ as a kind  of \textbf{universal model} of both classical and quantum states replacing the standard notion of (classical or quantum) statistical model of information geometry.
%%%%%%%%%%%%%%%
Objects in $\mathsf{NCP}$ would replace points in the manifold underlying a statistical model, while the morphisms in $\mathsf{NCP}$ would be a kind of point-wise symmetry transformations.
%%%%%%%%%%%%%%%%%%%%%%%%%%%%%%%%%%%%%%
Similarly to how every point $m$ in a statistical model with underlying smooth manifold $M$ carries a tangent space $T_mM$ and a cotangent space $T_m^*M$, an object $(\mathscr{A},\rho)$ in $ \mathsf{NCP}$ carries the Gelfand–Naimark–Segal (GNS) Hilbert space $\mathcal{H}_\rho$\footnote{Recall that the GNS Hilbert space associated to a state $\rho$ is given by the completion of $\mathscr{A}/\mathscr{N}_\rho$ with respect to the pre-Hilbert structure $\langle\mathbf{x}\mid \mathbf{y}\rangle_{\rho}:=\rho(\mathbf{x}^{\dagger}\mathbf{y})$ where, $\mathbf{x}, \mathbf{y} \in \mathscr{A}$  and $N_{\rho}=\{\mathbf{x}\in\mathscr{A}\;|\;\;\rho(\mathbf{x}^{\dagger}\mathbf{x})=0\}$ is the so-called Gel'fand ideal.} at each object $(\mathscr{A},\rho)$.
%%%%%%%%%%%%%%%%%%%%%%%%%%%%%%
The assignment $(\mathscr{A},\rho)\mapsto\mathcal{H}_\rho$ leads to a contravariant functor\footnote{Here, $\mathsf{Hilb}$ denotes the category of complex Hilbert spaces with bounded linear contractions.} $\mathfrak{G}: \mathsf{NCP} \to \mathsf{Hilb}$ defined on objects by $\mathfrak{G}_0(\mathscr{A},\rho)=\mathcal{H}_\rho$ and on morphisms by $\mathfrak{G}_1(\Phi) = \widetilde{\Phi}$ where, $\widetilde{\Phi}$ is the contraction determined by the CPU map $\Phi$. 
%%%%%%%%%%%%%%%%%%%%%%%%%%%%%%%%%%%%%
We call $\mathfrak{G}$ the GNS functor.
%%%%%%%%%%%%%%%%%%%%%%%%%%%%%%%%%%%%%

A Riemannian metric tensor $\mathtt{R}$ on a finite-dimensional manifold $M$ can be seen as the (smooth) assignment of a real Hilbert structure on the tangent space $T_mM$ at each $m\in M$.
%%%%%%%%%%%%%%%%%%%%%%%%%%%%%%%%%%%%%
Such an assignment can be seen as a covariant functor\footnote{Here, $\mathsf{Hilb}_{\mathbb{R}}$ denotes the category of real Hilbert spaces with bounded linear contractions.}  $\mathfrak{R}:\mathsf{C(M)}\to\mathsf{Hilb}_{\mathbb{R}}$ where, $\mathsf{C(M)}$
is the category associated with the manifold $M$, the category whose objects are points of $M$, and whose morphisms are the identity morphisms at each object.
%%%%%%%%%%%%%%%%%%%%%%%%%%%%%%%%%%%%%
Note that, as it is defined, the category $\mathsf{C(M)}$ does not recover any topological or differentiable properties of $M$.
%%%%%%%%%%%%%%%%%%%%%%%%%%%%

The functor $\mathfrak{R}$ is defined on objects by $\mathfrak{R}_0(m) := (T_mM, \mathtt{R}_m)$ and on morphisms by $\mathfrak{R}_1(1_m) := T(1_m): (T_mM, \mathtt{R}_m)\to(T_mM, \mathtt{R}_m)$.\footnote{Of course, a suitable condition realizing the notion of smoothness is also required, but for practical reasons we will not focus on it.}
%%%%%%%%%%%%%%%%%%%%%%%%%%%%%%%%%%%%
Moreover, when a Lie group $G$ acts on a manifold $M$, we have the action Lie groupoid $G\ltimes M$ \cite{Mackenzie-2005}, and a $G$-invariant Riemannian metric tensor $\mathtt{R}$ on $M$ gives rise to the functor $\mathfrak{R}: G\ltimes M\to\mathsf{Hilb}_{\mathbb{R}}$ defined on objects by $\mathfrak{R}_0(m) := (T_mM, \mathtt{R}_m)$, and on morphism by $\mathfrak{R}_1(\alpha\equiv(\mathrm{g},m)) :=  T\alpha: (T_mM, \mathtt{R}_m)\to(T_{\mathrm{g}\circ m}M, \mathtt{R}_{\mathrm{g}\circ m})$.
%%%%%%%%%%%%%%%%%%
Note that the invariance behaviour of $\mathtt{R}$ is encoded in the functoriality of $\mathfrak{R}$.
%%%%%%%%%%%%%%%%%%%%%%%%%%%%%%%
If we replace Riemannian metric tensors with their contravariant counterparts defined on cotangent spaces, we obtain contravariant functors.
%%%%%%%%%%%%%%%%%%%%%%%%%%%%%%%

Considering the analogies above, in the categorical setting of $\mathsf{NCP}$, we may replace a contravariant Riemannian metric tensor with a contravariant functor $\mathfrak{C}$ from $\mathsf{NCP}$ to $\mathsf{Hilb}$.
%%%%%%%%%%%%%%%%%%%%%%%%%%%%%%%%%%%%%%%
On objects of $\mathsf{NCP}$, the functor $\mathfrak{C}$ reads $\mathfrak{C}_0(\mathscr{A},\rho) = (\mathcal{K}_\rho, \mathfrak{C}_\rho)$ where, $\mathcal{K}_\rho$ is a Hilbert space and $\mathfrak{C}_\rho$ its Hilbert product, and on morphisms, it reads $\mathfrak{C}_1(\Phi: (\mathscr{A},\rho)\to(\mathscr{B},\sigma)) = \hat{\Phi}$, with $\hat{\Phi}\colon \mathcal{K}_\sigma \rightarrow\mathcal{K}_\rho$ a linear contraction.
%%%%%%%%%%%%%%%%%%%%%%%%%%%%%%%%%%%%%%%%%%%%%%%%%%%%%%
Taking into account that all contravariant Riemannian metric tensors on $M$ are defined on the same cotangent space, we impose the strong equality
\begin{equation} \label{eqn: categorical monotonicity 1}
\mathfrak{F}\circ \mathfrak{C} = \mathfrak{F}\circ \mathfrak{G},	
\end{equation}  
where $\mathfrak{F}\colon\mathsf{Hilb}\rightarrow\mathsf{Vect}$ is the obvious forgetful functor.
%%%%%%%%%%%%%%%%%
%Moreover, since this pre-Hilbert structure should satisfy some symmetry constraints we may described it by the contravariant functor $\mathfrak{C}\colon \mathsf{NCP} \rightarrow\mathsf{Hilb}$ satisfying the condition
%\begin{equation} \label{eqn: categorical monotonicity 1}
%\mathfrak{F}\circ \mathfrak{C} = \mathfrak{F}\circ \mathfrak{G},	
%\end{equation}  
%where, $\mathfrak{F}\colon\mathsf{Hilb}\rightarrow\mathsf{Vect}$ is the obvious forgetful functor.
%%%%%%%%%%%%%%%%%%%%%%%%%%%%%%%%%%%%%%%%
The previous condition guarantees that the vector space underlying the Hilbert space $\mathfrak{C}_{0}(\mathscr{A},\rho)=\mathcal{K}_\rho$ is the vector space underlying the GNS Hilbert space $\mathcal{H}_{\rho}$, and that the morphism $\mathfrak{C}_{1}(\Phi)$ is the linear contraction $\tilde{\Phi}$ determined by the CPU map $\phi$ inducing the morphism $\Phi$ in $\mathsf{NCP}$.
%%%%%%%%%%%%%%%%%%%%%%%%%%%%%%%
We will call such a contravariant functor $\mathfrak{C}$ a \textit{field of covariances} on $\mathsf{NCP}$.
%%%%%%%%%%%%%%%%%%%%%%%%%%%%%%%%%%%%
The alternative Hilbert product $\mathfrak{C}_\rho$, defined on $\mathcal{H}_\rho$ via the functor $\mathfrak{C}$, will be referred to as a covariance at $\rho$.
%%%%%%%%%%%%%%%%%%%%%%%%%%%%%%%%%%
%However, in order for this analogy to be well posed a suitable notion of ``smooth'' assignment should be defined for field of covariances.
%%%%%%%%%%%%%%%%%%%%%%%%%%
%Such notion will be defined by exploiting the weak*-topology on the space of states\footnote{This and other technical details will be explain in a forthcoming work.}.  
%%%%%%%%%%%%%%%%%%%%%%%%%%%%%%%%%

This  perspective, which favors contravariant Riemannian metric tensors over their covariant counterparts aligns with the original derivation of the Cramér-Rao bound which is expressed in terms of  a contravariant Riemannian metric \cite{Hendriks-1991,Nagaoka-2024}.
%%%%%%%%%%%%%%%%%%%%%%%%%%%%%%%%%%%
Moreover, the term ``covariance'' is motivated by the fact that, when the algebra under investigation is $\mathscr{A}\cong\mathcal{L}^\infty(\Omega,\nu)$ and the state $\rho$ is normal (which means it is associated with a probability measure on $\Omega$ which is absolutely continuous with respect to $\nu$), the  GNS  Hilbert space is $\mathcal{H}_\rho\cong\mathcal{L}^2(\Omega,\rho)$, and the statistical covariance between (complex-valued) random variables (with vanishing expectation with respect to $\rho$)  coincides with their scalar product with respect to the GNS inner product.
%%%%%%%%%%%%%%%%%%%%%%%%%%%%%%%%%%%%

In this categorical setting, the analogue of Čencov's problem would be to classify all functors $\mathfrak{C}:\mathsf{NCP}\to\mathsf{Hilb}$ satisfying condition \eqref{eqn: categorical monotonicity 1}.
%%%%%%%%%%%%%%%%
From the point of view of covariances, the functoriality of $\mathfrak{C}$ on morphisms encodes the monotonicity property 
\begin{equation}\label{eqn: categorical monotonicity 3}
\mathfrak{C}_{\rho}(\widetilde{\Phi}(\xi),\widetilde{\Phi}(\xi))\leq \mathfrak{C}_{\sigma}(\xi,\xi),
\end{equation}
for any $\Phi\in\mathsf{NCP}_{1}$. 
%%%%%%%%%%%%%%%%%%%%%%%%%%%%%%%%%

It is worth noting that the GNS functor is an obvious example of field of covariance on $\mathsf{NCP}$, that reduces to the (complexified) statistical covariance as mentioned above.
%%%%%%%%%%%%%%%%%%%%%%%%%%%%%%%
Since the Fisher-Rao metric tensor is essentially the inverse of the statistical covariance, this instance suggests a strong connection between the GNS  functor and the Fisher-Rao metric tensor that will be explored in a forthcoming publication.
%%%%%%%%%%%%%%%%%%%%%%5

%%%%%%%%%%%%%%%
We will address the problem of classification of fields of covariances on $\mathsf{NCP}$ in a series of forthcoming publications.
%%%%%%%%%%%%%%%%%%%%%%%%%%%%%%%%%
In particular, it is worth mentioning a result of our investigation.
%%%%%%%%%%%%%%%%%%%%
If we focus on the full subcategory $\mathsf{fNCT}$ of tracial states\footnote{A state $\tau$ on $\mathscr{A}$ is called \textit{tracial} if $\tau(ab)=\tau(ba)$ for every $a,b\in\mathscr{A}$.} on finite-dimensional algebras, a uniqueness result reminiscent of Čencov's classical result holds because the GNS functor $\mathfrak{G}$ is the only possible functor (up to the choice of two positive constants).
%%%%%%%%%%%%%%%%%%%%
This case is particularly interesting because it shows that uniqueness is possible also in a non-commutative setting, as long as the states under consideration do not feel the non-commutative character of the algebra.
%%%%%%%%%%%%%%%%%%%%%%%%%%%

%However, it is important to highlight that focusing on suitable subcategories of $\mathsf{NCP}$ we are able to obtain classification problems that are very close to that of Čencov and Petz. 
%%%%%%%%%%%%%%%%%%%%%%%%%%%%%%%%%
%In particular, the study of finite-dimensional W*-algebras reveals an ``intermediate'' case between Čencov's and Petz's frameworks as we found that the uniqueness of the Fisher metric stems not from the commutativity of the algebra  but rather from the nature of the states.
%%%%%%%%%%%%%%%%%%%%%%%%%%%%%%%%%

\section{Statistical models in $\mathsf{NCP}$}\label{sec: section on statistical models}

As remarked in section \ref{sec: section on covariances},  we may intuitively think of $\mathsf{NCP}$ as a sort of universal statistical model that encompasses all classical and quantum statistical models of information geometry. 
%%%%%%%%%%%%%%%%%%%%%%%%%%%%%%%%%%%%%%%%%%%%%%%%%%
This idea can be further formalized by viewing models as subcategories of $\mathsf{NCP}$. 
%%%%%%%%%%%%%%%%%%%%%%%%%%%%%%%%%%%%%%%%%%%%%%%%%%
Specifically, let $M$ denote the manifold underlying a statistical model, and let $\mathsf{C(M)}$ be its associated category. 
%%%%%%%%%%%%%%%%%%%%%%%%%%%%%%%%%%%%%%%%%%%%%%%%%%
We can define the trivial immersion functor $\hat{\mathtt{i}}: \mathsf{C(M)} \to\mathsf{NCP}$ by setting $\hat{\mathtt{i}}_0(m):= (\mathscr{A},\rho_m)$, where $\hat{\mathtt{i}}_0(m) \cong \hat{\mathtt{i}}_0(n) \iff m\cong n$, and $\hat{\mathtt{i}}_1(1_m) := 1_{{{\mathtt{i}}_0}(m)}$.
%%%%%%%%%%%%%%%%%%%%%%%%%%%%%%%%%%%%%%%%%%%%%%%%%%%%
The image of this functor is trivially a subcategory of $\mathsf{NCP}$, and any statistical model can be seen as a (trivial) subcategory of $\mathsf{NCP}$.  
%%%%%%%%%%%%%%%%%%%%%%%%%%%%%%%%%%%%%%%%%%%%%%%%%%

Building on the manifold-oriented perspective of information geometry, we may consider the map on objects $\hat{\mathtt{i}}_0: M \to \mathsf{NCP}_0, \, \hat{\mathtt{i}}_0(m):= (\mathscr{A},\rho_m)$ as the analogue of the injective map $i: M \to \mathcal{S}(\Omega)$ in the classical case or $i: M \to (\mathcal{B(H)})^*$ in the quantum case, which is at the core of the definition of a statistical model \cite{Amari-2016}. 
%%%%%%%%%%%%%%%%%%%%%%%%%%%%%%%%%%%%%%%
In this context, the introduction of an essentially injective functor, which is crucial for obtaining a subcategory, can be understood as the categorical analogue of the injectivity of the immersion map.
%%%%%%%%%%%%%%%%%%%%%%%%%%%%%%%%%%%%%%%%
Therefore, in the categorical context of $\mathsf{NCP}$, it is natural to define a statistical model as a subcategory of $\mathsf{NCP}$, referred to as a \textit{statistical subcategory}.
%%%%%%%%%%%%%%%%%%%%%%%%%%%%%%%%%%%%%%%

The immersion functor $\hat{\mathtt{i}}$ defined above is not able to encode additional geometric structures of the model under investigation, but the categorical framework is perfectly capable of handling additional structures.
%%%%%%%%%%%%%%%%%%%%%%%%%%%%%%%%
%one that may allow us to bring this additional structure into the picture
For instance, let us examine the case of univariate normal distributions.
%%%%%%%%%%%%%%%%%%%
The manifold underlying the model of univariate normal distributions is $M=\mathbb{R}\times\mathbb{R}^{+}$, which admits a Lie group structure.
%%%%%%%%%%%%%%%%%%
Moreover, the ``same" Lie group can be thought of as the affine group naturally acting the outcome space $\Omega=\mathbb{R}$ on which univariate normal distributions are defined.
%%%%%%%%%%%%%%%%%%%%%%%%%%
%meaning the manifold of parameters can be endowed with a natural Lie group structure, which allows the definition of an action over the outcome space, which corresponds to the affine action on $\mathbb{R}$  \cite{A-N-2000}.
%%%%%%%%%%%%%%%%%%%%%%%%%%%%%%%%
Such a geometric structure can be encoded within $\mathsf{NCP}$.  
%%%%%%%%%%%%%%%%%%%%%%%%%%%%%%%%%
Concretely, given the affine action $\alpha_\xi: \mathbb{R}\to\mathbb{R}, x\mapsto \sigma x + \mu$ with $\xi \equiv (\mu,\sigma) \in M $,  we may define the automorphism 
\begin{equation}\label{eqn: automorphism for the case of Gaussians}
\phi_\xi:\mathcal{L}^\infty(\mathbb{R},dl)\to\mathcal{L}^\infty(\mathbb{R},dl), f \mapsto \phi_\xi(f) := f \circ \alpha_\xi
\end{equation}
where, $l$ is the Lebesgue measure on $\mathbb{R}$.
%%%%%%%%%%%%%%%%%%%%%%%%%%%%%%%%%
Moreover, $\Phi_\xi^*(p_{\xi'}) = p_{\xi \circ \xi'}$ which can be expressed as 
$$\int_A p(\xi',x)dl(x) = \int_{\xi\circ A} p(x, \xi\circ\xi')dl(x)$$
ensuring the compatibility between the Lie group structure of  $ M \cong \mathbb{R}\times\mathbb{R}^+$ and the realization of its points as probability distributions on $\mathbb{R}$ \cite{A-N-2000}.
%%%%%%%%%%%%%%%%%%%%%%%%%%%%%%%%%
Thus, the action of $M$ over itself defines an action groupoid\footnote{Recall that every groupoid is a category.} which can be embedded as a subcategory of $\mathsf{NCP}$ by means of the image of the Gaussian functor $\hat{\mathtt{G}}: M\ltimes M\to\mathsf{NCP}$, $\hat{\mathtt{G}}_0(\xi'):= (\mathcal{L}^\infty(\mathbb{R},dl),p_{\xi'})$, where $\hat{\mathtt{G}}_0(\xi') \cong \hat{\mathtt{G}}_0(\lambda) \iff \xi'\cong\lambda$, and $\hat{\mathtt{G}}_1((\xi,\xi')) := \Phi_\xi: (\mathcal{L}^\infty(\mathbb{R},dl),p_\xi)\to(\mathcal{L}^\infty(\mathbb{R},dl),p_{\xi\circ\xi'}) $ with $\phi_\xi$ given by equation \eqref{eqn: automorphism for the case of Gaussians}.
%%%%%%%%%%%%%%%%%%%%%%%%%%%%%%%%%
From this particular example, a specific structure can be distilled for statistical models admitting a Lie group action.
%that admit a group structure (not necessarily Lie).
%%%%%%%%%%%%%%%%%%%%%%%%%%%%%%%%%%
Specifically, the Lie group action is encoded in  the  functor $\hat{\mathtt{g}}: G\ltimes M\to\mathsf{NCP}$, 
$\hat{\mathtt{g}}_0(m):= (\mathscr{A},\rho_m)$, where $\hat{\mathtt{g}}_0(m) \cong \hat{\mathtt{g}}_0(n) \iff m\cong n$, and $\hat{\mathtt{g}}_1((g,m)) := \Phi_g$ where $\phi_g: \mathscr{A}\to\mathscr{A}$ is an automorphism of $\mathscr{A}$ and $\phi_g^*(\rho_m) = \rho_{g\circ m}$ whose image defines a subcategory of $\mathsf{NCP}$.
%%%%%%%%%%%%%%%%%

Following this line of reasoning, it is a matter of verification that transformation models and coherent states are statistical subcategories of $\mathsf{NCP}$. 
%%%%%%%%%%%%%%%%%%%%%%%%%%%%%%%%%
Moreover, other interesting statistical subcategories arise in this way once we relax the requirement of dealing with smooth manifolds.
%%%%%%%%%%%%%%%%%%%%%%
For example, we can take topological action groupoid determined by the topological space  $\mathcal{S}(\mathscr{A})$ of states on the $C^*$-algebra $\mathscr{A}$ with the natural action of the automorphism group $\mathrm{Aut}(\mathscr{A})$, and realize it as a statistical subcategory of $\mathsf{NCP}$ by means of the obvious functor given by identification.
%%%%%%%%%%%%%%%%
These statistical subcategories contain subcategories that are meaningful in their own right(for instance, when $\mathscr{A}=\mathcal{B}(\mathcal{H})$, the action groupoid of pure states with the unitary group is an example, while, imposing $\mathrm{dim}(\mathcal{H})<\infty$ allows us to recover the action groupoids of the manifolds of quantum states with fixed rank with the action of the unitary group).
%%%%%%%%%%%%%%%%%%%%%%%

%we may consider embedding the action groupoid given by the adjoint action of the group of automorphisms of a W*-algebra over the set of normal states. 
%%%%%%%%%%%%%%%%%%%%%%%%%%%%%%%%%%%%%%%
%This groupoid is not a Lie groupoid in general; however, when the algebra is finite-dimensional and we consider manifolds of states of fixed rank, we obtain interesting groupoids.
%%%%%%%%%%%%%%%%%%%%%%%%%%%%%%%%%%%%%%%%
%The embedding of these groupoids gives rise, for example, to pure states, faithful states, and coherent states in $\mathsf{NCP}$.
%%%%%%%%%%%%%%%%%%%%%%%%%%%%%%%%%%%%%
%Furthermore, when the group of automorphisms is discrete—such as in the case where the algebra is $\mathscr{A}\cong\mathbb{C}^n$—such an embedding leads to probability vectors.
%%%%%%%%%%%%%%%%%%%%%%%%%%%%%%%%%%%%%%%%%
The realization of models that do not admit a Lie group action as non-trivial statistical subcategories of $\mathsf{NCP}$, will be discussed in a forthcoming work.
%%%%%%%%%%%%%%%%%%%%%%%%%%%%%%%%%%%%%%%%

From the discussion above, it follows that the notion of statistical subcategories of $\mathsf{NCP}$ allows us to connect with and generalize the notion of (classical and quantum) statistical models. 
%%%%%%%%%%%%%%%%%%%%%%%%%%%%%%%%%%%%%%%%%%%%%%%%%%
The next step would be that of using a field of covariance on $\mathsf{NCP}$ to induce a Riemannian metric tensor on a statistical subcategory admitting a smooth structure in a way that suitably generalizes the idea of taking the pullback of covariant tensor fields.
%%%%%%%%%%%%%%%%%%%%%
In this way, the notion of field of covariance would act as a sort of universal model for the Riemannian geometries of statistical models.
%%%%%%%%%%%%%%%%%%%%%%%%%%%%%%%%%%%
We plan to thoroughly discuss this procedure in a forthcoming work, showing how the GNS functor discussed in section \ref{sec: section on covariances} leads to the Fisher-Rao metric tensor on statistical subcategories on commutative algebras, and to the Bures-Helstrom metric tensor on the statistical subcategory of faithful states on $\mathcal{B}(\mathcal{H})$ in finite dimensions.

\addcontentsline{toc}{section}{References}
\bibliographystyle{plain}
\bibliography{scientific_bibliography}

\end{document}